\documentclass[oneside]{article}
\usepackage[T1]{fontenc}
\usepackage{authblk}
\usepackage{float}
\usepackage{amsmath}
\usepackage{graphicx}
\usepackage{xfrac}
\usepackage[english]{babel}
\usepackage{lmodern}
\usepackage[T1]{fontenc}
\usepackage[latin9]{inputenc}

\usepackage{csquotes}

\usepackage{mathtools}
\usepackage{graphicx}
\usepackage{esint}
\usepackage[unicode=true,
 bookmarks=false,
breaklinks=false,pdfborder={0 0 1},backref=false,colorlinks=false]
 {hyperref}
\usepackage[a4paper, total={210mm,297mm},margin=2.0cm]{geometry}

\title{Modelling pattern formation in soft flowing crystals}

\author[1]{Andrea Montessori \thanks{Electronic address: \texttt{and.montessori@gmail.com}; Corresponding author}}

\author[1]{Marco Lauricella}

\author[2]{Adriano Tiribocchi}

\author[2,1]{Sauro Succi}

\affil[1]{Istituto per le Applicazioni del Calcolo CNR, via dei Taurini 19, Rome, Italy}
\affil[2]{Center for Life Nano Science@La Sapienza, Istituto Italiano di Tecnologia, 00161 Roma, Italy}

\date{\today}

\begin{document}

\maketitle

\begin{abstract}
We present a mesoscale representation of near-contact interactions between colliding droplets which
permits to reach up to the scale of full microfluidic devices, where such droplets are produced.
The method is demonstrated for the case of colliding droplets and the formation
of soft flowing crystals in flow-focussing microfluidic devices.
This model may open up the possibility of multiscale simulation of microfluidic 
devices for the production of new droplet/bubble-based mesoscale porous materials.
\end{abstract}

\section{Introduction}

In the recent years, major advances in material science, fuelled by fundamental progress in condensed matter physics, have lead to important technological developments in many fields of science and engineering.

In particular, condensed matter physics has made substantial strides in the field  of soft matter, namely the study of complex states of matter, typically composed of polymers, colloids, surfactants, liquid crystals, and other mesoscopic constituents, whose behaviour is dictated by energies at the thermal scale, i.e. ``$kT$'' physics.

Many soft materials exhibit physical and rheological properties  which cannot be inferred directly from those of their constitutive elements \cite{tiribocchi2019curvature,gai2019timescale,sollich1997rheology,goyon2008spatial}. For instance, foams, binary mixtures of gas and liquid, display mechanical properties which cannot be traced back to either of the two \cite{dollet2015two,gai2016spatiotemporal}.  

In this respect, progresses in microfluidic technologies have lead to a precise control over the production of bubbles and droplets in micro-channels, which can represent the fundamental constituents for the production of novel soft mesoscale materials \cite{li2018microfluidic,stolovicki2018throughput,montessori2018elucidating,montessori2019jetting}.

This has stimulated an outburst of experimental and theoretical activity alike, including computational methods, aiming at shedding light into the basic mechanisms controlling the behaviour of soft matter systems \cite{Murtola2009pccp,falcucci2011lattice,allen2017computer}.

In particular, a wide body of theoretical and experimental work has elucidated the complex nature of the interactions occurring within intervening liquid films (i.e. near-contact interactions) \cite{derjaguin1940,derjaguin1941,verwey1948,webber2008,dagastine2006,chan2011film,wang2015,montessori2019mesoscale}. This layed down the foundations for describing a broad variety of complex systems, ranging from colloids, foams and emulsions, to flowing collections of droplets and bubbles characterized by highly ordered and uniform crystal-like structures, known as \textit{soft flowing crystals} \cite{garstecki2006flowing,marmottantprl2009}.

Computational methods often permit to investigate parameter regions not accessible to experiments, and to explore non-perturbative regimes out of reach to analytical methods. However, the direct introduction of the near-contact forces is computationally prohibitive, as it requires to simultaneously solve almost six spatial decades: from millimeters, i.e. the typical size of microfluidic devices, all the way down to nanometers, namely the relevant spatial scale of contact forces \cite{montessori2019mesoscale}. 

Alternatively, a ``softer'' path can be taken. This does not require the direct introduction of nanoscale forces at the interface level, but relies on suitable coarse-grained models incorporating effective forces and potentials designed in such a way to retain the essential physics of soft interactions. While inherently approximate in nature, the latter alternative is very appealing, as it offers dramatic computational savings.

The success of such  strategy is strictly dependent on the degree of universality of the underlying physics. Universality means dependence on suitable dimensionless groups rather than on the specific details of competing forces and interactions \cite{succi2015lattice}.

In this paper we show that a multicomponent lattice Boltzmann (LB) model \cite{higuera1989lattice,montessori2015lattice,montessori2019jfm,montessori2019mesoscale,montessori2018regularized}, augmented with a suitable forcing term aimed at representing the effects of the near-contact forces operating at the fluid interface level \cite{Benzi2009,Sbragaglia2012,Fei2018}, is capable of reproducing the droplets self-assembly and, for the first time, the morphology and self-tuning of the soft crystal patterns (from Hex-two to Hex-one structure) observed in  microfluidic channels, in good agreement with previous experimental works \cite{marmottantprl2009,marmottant2009}. 

\section{Method}

In this section, we provide a brief account of the numerical model, an extended color-gradient approach with repulsive near contact interactions, previously introduced in Ref. \cite{montessori2019jfm}. In the color gradient LB for multicomponent flows, two sets of distribution functions track the evolution of the two fluid components, according to the usual streaming-collision algorithm (see \cite{succi2018lattice,kruger2017lattice}):

\begin{equation} \label{CGLBE}
f_{i}^{k} \left(\vec{x}+\vec{c}_{i}\Delta t,\,t+\Delta t\right) =f_{i}^{k}\left(\vec{x},\,t\right)+\Omega_{i}^{k}[ f_{i}^{k}\left(\vec{x},\,t\right)]+S_i(\vec{x},t),
\end{equation}

where $f_{i}^{k}$ is the discrete distribution function, representing the probability of finding a particle of the $k-th$ component at position $\vec{x}$, time $t$ with discrete velocity $\vec{c}_{i}$, and $S_i$ is a source term coding for the effect of external forces (such as gravity, near-contact interactions, etc). For further details about the LB equation see \cite{succi2018lattice,kruger2017lattice}. 
\begin{figure}
\begin{center}
\includegraphics[scale=0.2]{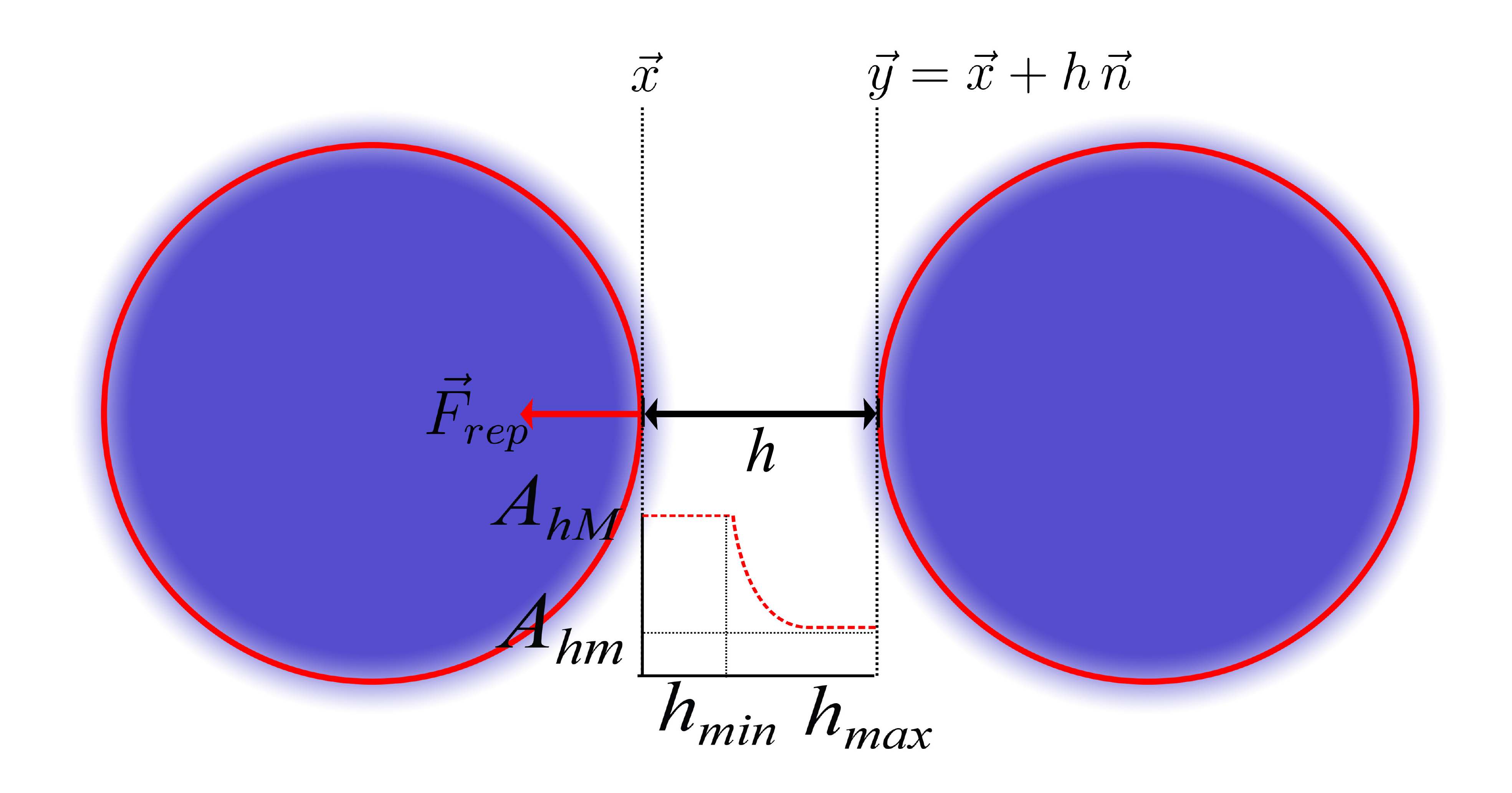}
\caption{Near contact forces representation. Mesoscale modelling of near contact interactions between two immiscible fluid droplets. $\vec{ F}_{rep}$ represents the repulsive force and $\vec{ n}$ is the unit vector perpendicular to the fluid interface. The vectors $\vec{x}$ and $\vec{y}$ indicate the positions, placed a distance $h$, taken within the fluid interface and the dotted line tracks its outermost frontier.
$A_{hM}$ and $A_{hm}$ are, respectively, the maximum and minimum value of the repulsive parameter.}
\label{fig0}
\end{center}
\end{figure}
The lattice time step is taken equal to $1$, and the index $i$ runs over the discrete lattice
directions $i = 1,...,b$, where $b=27$ for a three dimensional twenty-seven speed lattice (D3Q27).
The density $\rho^{k}$ of the $k-th$ component and the total momentum of the mixture 
$\rho \vec{u}=\sum_k\rho^{k}\vec{u^k} $  are given by the zeroth and the first order moment 
of the distribution functions
$\rho^{k}\left(\vec{x},\,t\right) = \sum_i f_{i}^{k}\left(\vec{x},\,t\right)$ and 
$\rho \vec{u} = \sum_i  \sum_k f_{i}^{k}\left(\vec{x},\,t\right) \vec{c}_{i}
$.
The collision operator splits into three components \cite{gunstensen1991lattice,leclaire2012numerical,leclaire2017generalized}: 

\begin{equation}
\Omega_{i}^{k} = \left(\Omega_{i}^{k}\right)^{(3)}\left[\left(\Omega_{i}^{k}\right)^{(1)}+\left(\Omega_{i}^{k}\right)^{(2)}\right].
\end{equation}

In the above, $\left(\Omega_{i}^{k}\right)^{(1)}$, stands for the standard collisional relaxation \cite{succi2018lattice}, 
$\left(\Omega_{i}^{k}\right)^{(2)}$ is the perturbation step \cite{gunstensen1991lattice}, which contributes to the buildup of the interfacial tension. 
Finally, $\left(\Omega_{i}^{k}\right)^{(3)}$ is the recoloring step \cite{gunstensen1991lattice,latva2005diffusion}, which promotes 
the segregation between the two species, so as to minimise their mutual diffusion.

By performing a Chapman-Enskog expansion, it can be shown that the hydrodynamic limit of Eq.\ref{CGLBE} converges to a
set of equations for the conservation of mass and linear momentum, with a capillary stress tensor of the form:
\begin{equation}
\boldmath{\Sigma}=-\tau\sum_i \sum_k\left(\Omega_{i}^{k}\right)^{(2)} \vec{c}_i \vec{c_i}= \frac{\sigma}{2 |\nabla \rho|}(|\nabla \rho|^2\mathbf{I} - \nabla \rho \otimes \nabla \rho)
\end{equation}
being $\tau$ the collision relaxation time, controlling the kinematic viscosity via the relation  
$\nu=c_s^2(\tau-1/2)$ ( $c_s=1/\sqrt{3}$ the sound speed of the model), $\sigma$ is the surface 
tension \cite{succi2018lattice,kruger2017lattice} and $\otimes$ denotes a dyadic tensor product. 

The stress-jump condition across a fluid interface is augmented with a repulsive term aimed at 
providing a mesoscale representation of all the repulsive near-contact forces (i.e., Van der Waals, electrostatic, steric 
and hydration repulsion) acting on much smaller scales ($\sim  O(1 \; nm)$)  than those resolved 
on the lattice (typically well above hundreds of nanometers).

It takes the following form:
\begin{equation}
\mathbf{T}^1\cdot \vec{n} - \mathbf{T}^2 \cdot \vec{n}=-\nabla(\sigma \mathbf{I} - \sigma (\vec{n}\otimes \vec{n})) - \pi \vec{n}
\end{equation}
where $\pi[h(\vec{x})]$ is responsible for the repulsion between neighboring fluid interfaces, 
$h(\vec{x})$ being the distance along the normal $\vec{n}$, between locations 
$\vec{x}$ and $\vec{y}=\vec{x}+ h \vec{n}$ at the two interfaces, respectively (see Fig. \ref{fig0}).\\

The above expression can be recast in the following form \cite{li2016macroscopic}:
\begin{equation}
\mathbf{T}^1\cdot \vec{n} - \mathbf{T}^2 \cdot \vec{n}=\sigma (\nabla \cdot \vec{n})\vec{n} - \nabla_t\sigma - \pi \vec{n}
\end{equation}
in which $\nabla_t$ identifiesy the gradient tangent to the interface.

By neglecting any variation of the surface tension along the interface, we can approximate  
$\mathbf{T}=-p\mathbf{I}$ \cite{brackbill1992continuum} and the above equation takes the following form:
\begin{equation}
(-p_1\mathbf{I}) \cdot \vec{n}- (-p_2\mathbf{I})\cdot \vec{n}=\sigma (\nabla \cdot \vec{n})\vec{n}  - \pi \vec{n}
\end{equation}

By projecting the equation along the normal to the surface, we obtain the  \textit{extended} Young-Laplace 
equation \cite{chan2011film, williams1982nonlinear}:
\begin{equation}
(p_2 - p_1)=\sigma (\nabla \cdot \vec{n}) - \pi
\end{equation}

The additional term can be readily included within the LB framework, by adding a 
forcing term acting only on the fluid interfaces in near contact, namely:
\begin{equation}
\vec{F}_{rep}= \nabla \pi := - A_{h}[h(\vec{x})]\vec{n} \delta_I
\end{equation}

In the above, $A_h[ h(\vec{x})]$ is the parameter controlling the strength (force per unit volume)
of the near contact interactions, and is set equal to a constant $A_{hM}$ if $h<h_{min}$, whereas it decreases as $h^{-3}$ if $h>h_{min}$. The term $h_{min}$ sets the threshold below which the repulsive force is effective. In our simulations we keep $h_{min}$ equal to four lattice sites, and $A_{hM}=0.05$. In addition, $h(\vec{x})$ is the distance between the interfaces, $\vec{n}$ is a unit vector normal to the interface and $\delta_I\propto \nabla\phi$ is a function, proportional to the phase field $\phi=\frac{\rho^1-\rho^2}{\rho^1+\rho^2}$, that localizes the force at the interface.

The further repulsive force (added to the right hand side of Eq.~\ref{CGLBE}) naturally leads to the following (extended) 
conservation law for the momentum equation:
\begin{equation} \label{NSEmod}
\frac{\partial \rho \vec{u}}{\partial t} + \nabla \cdot {\rho \vec{u}\vec{u}}=-\nabla p + \nabla \cdot [\rho \nu (\nabla \vec{u} + \nabla \vec{u}^T)] + \nabla \cdot (\boldmath{\Sigma}  +   \pi \mathbf{I})
\end{equation}
This is the Navier-Stokes equation for a multicomponent system, augmented with a surface-localized repulsive 
term, expressed through the gradient of the potential function $ \pi$.

\section{Results}

In the following, we demonstrate the proposed scheme for two relevant applications, the 
collision of immiscible liquid droplets and the formation of flowing crystals in microfluidic devices. 

\subsection{Colliding droplets}

We first show the capability of the extended LB model to accurately reproduce the correct dynamics of off-axis 
collisions between two immiscible droplets \cite{Chen2006}.
The characteristic non-dimensional parameters, governing the collision outcome are the Weber and the Reynolds 
numbers, defined as $We = \rho U^2_{rel} D/\sigma$ and
$Re = U_{rel}D/\nu$, respectively.
In the above, $U_{rel}$ the relative impact velocity, $D$ the droplet diameter, $\sigma$ the surface tension coefficient 
and $\nu$ the kinematic viscosity, as well as the impact number $b=\chi/D$, namely, the distance between the collision 
trajectories in units of the droplet diameter.

In this simulation the values of the physical and geometrical parameters are: domain size $121 \times 101 \times 121$ ($nx \times ny \times nz$), droplets diameter $D=30$, relative speed between the impacting droplets $U_{rel}=0.06$, surface tension $\sigma=0.01$, magnitude of the repulsive force $A_h=0.01$, kinematic viscosity $\nu=0.0167$, Weber number $We=10$, Reynolds number $Re=108$ and impact number $b=0.33$.

\begin{figure}
\begin{center}
\includegraphics[scale=1.5]{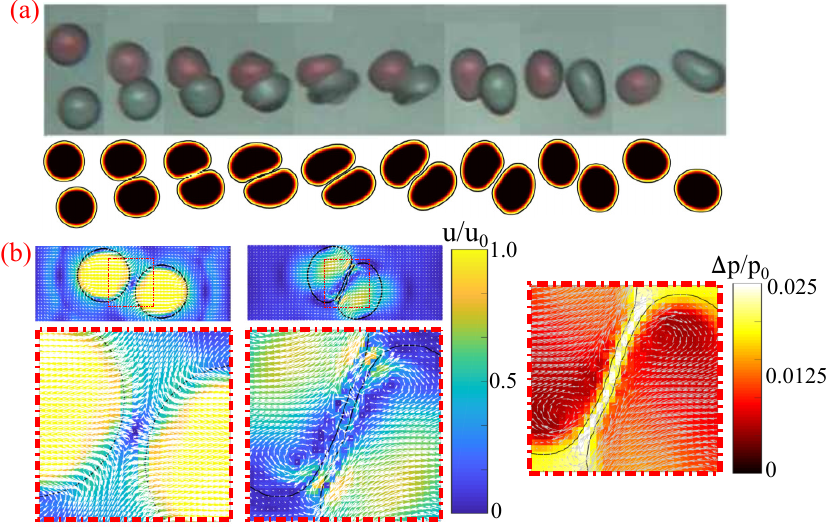}
\caption{\label{fig1} (a) Collision  sequence with  $b=0.33$ and $We=10$. The upper row shows the experiments of Ref.~\cite{Chen2006}, the lower row shows the simulation results. (b) Left. Sequence of the flow field during the impact between the two droplets (mid-plane slice). During the first stage of the collision the dispersed phase flows outwards, allowing the two droplets to approach. Afterwards, when the droplets come in close contact, the fluid within the thin film begins to recirculate, thereby preventing film rupture, hence the coalescence between the droplets. (b) Right. Map of the local pressure deviation $\Delta p/p_0=(p_0-p)/p_0$, where $p_0$ is the bulk pressure of the surrounding fluid and $p$ is the local pressure. As shown in the figure, the outer fluid is driven inwards and consequently stabilizes the film.}
\end{center}
\end{figure}

In figure \ref{fig1}(a) we report the collision sequence between the droplets and
compare the numerical results with the experimental data reported in \cite{Chen2006}.
The experiments were performed using liquid droplets with diameters ranging between $700-800 \mu m$ and 
impact velocities in the range of $1-3 m/s$. \\ 
As shown in panel (a) of figure \ref{fig1}, the two droplets undergo a kiss and tumble collision, in close agreement with the experiment. 

It is worth noting that the additional repulsive force proves instrumental to reproduce this collision outcome. 
Indeed, the coalescence between the droplets is frustrated by the effect of the near-contact repulsive forces, which 
prevent the rupture of the intervening thin film between the impacting droplets.\\
We then inspected the evolution of the thin liquid film during the collision process. 
As reported in fig. \ref{fig1}b (left), the fluid between the two approaching droplets first flows outwards,
and then recirculates within the intervening film,
producing a decrease in the fluid pressure (fig.~\ref{fig1}b,right) that temporarily
stabilizes the film and prevents the droplets' coalescence. After the collision, droplets are scattered far apart and the film gradually disappears.

The observed phenomenon resembles the so-called \textit{Marangoni flow} in liquid films, namely a fluid recirculation 
occurring in liquid thin films in the presence of shear and temperature gradients, which has been observed to prevent 
the coalescence between bodies of liquids \cite{dell1996suppression}.\\

\subsection{Crystal patterns in microchannels}

We then performed three-dimensional simulations of a oil/water emulsion in a microfluidic flow-focusing device \cite{link2004geometrically,garstecki2004formation}.
The micro-device is made of three inlet channels ($ H=200 \mu m$), supplying the dispersed  and the 
continuous  phase, an orifice placed downstream the three coaxial inlet streams ($h=100 \mu m$) plus an outlet channel of height $H_c=400 \mu m$
and length $1.9 \times 10^3 \mu m$.
The width of the focuser (the dimension perpendicular to the flow direction) is $W=100 \mu m$.
In our simulations we consider a device in which the thickness of the inlet and of the outlet channels is $L_x=20$ (perpendicular to the plane), while their height is $L_z=80$. Moreover, the length of the outlet channel is $L_y=420$, that of the inlet one is $L_y=40$, and the orifice is cubic shaped with dimensions $20\times 20\times 20$. Finally, we use no-slip conditions on both walls and, to achieve hydrophobicity, we set the contact angle with the dispersed phase at $\simeq 130^{o}$.

The mechanism of droplet formation follows from the periodic pinch-off of the dispersed jet by the continuous stream and the pinch-off mechanism takes place in the small orifice.

Such microfluidic geometry is widely employed for the production of foams and 
emulsions, as it offers an accurate control over the monodispersity and the droplet size 
\cite{whitesides2006origins, sackmann2014present, cruz2017jfm}.
The high degree of flow reproducibility is due to the dominance of the viscous forces over inertia, 
which smoothens the flow and tames hydrodynamic instabilities. 
The periodic pinch-off process permits to  produce
droplets with standard deviations in size as little as the $0.1\%$  \cite{marmottant2009microfluidics,ganan2001perfectly,link2004geometrically}.

Here, we show that the proposed approach is able to reproduce different crystal patterns in the outlet channel of the device.

By taking an interfacial tension of an oil-water mixture ($\sim 50 mN/m$), the dynamic viscosity of the water 
(dispersed phase) $\mu \sim 10^{-3} Pa\cdot s$ and an inlet velocity of the dispersed phase $\sim 0.1 m/s $, 
we obtain a Weber number $We=0.04$ and a capillary number $Ca=0.0017$.
\begin{figure}
\begin{center}
\includegraphics[scale=0.7]{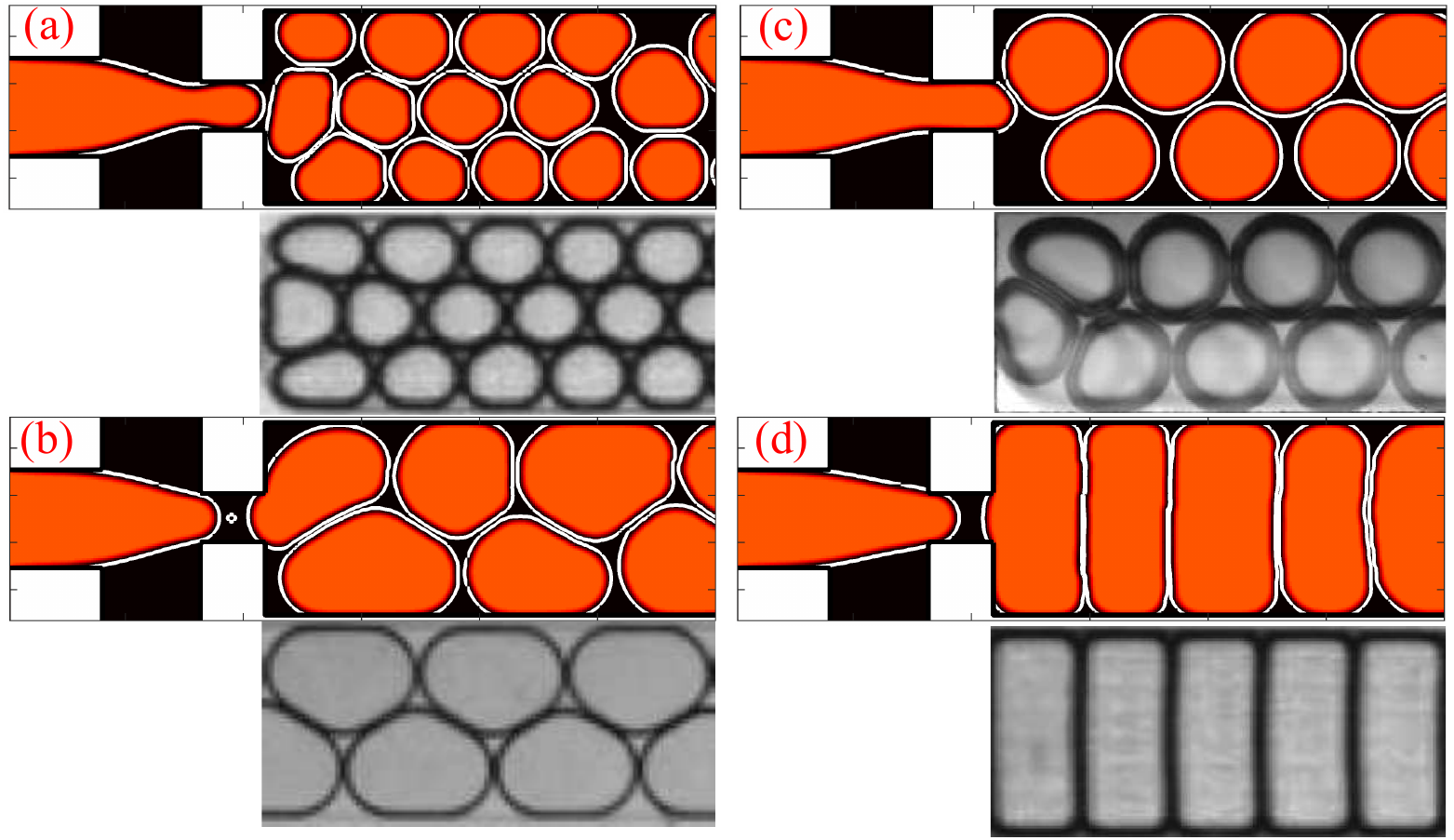}
\caption{\label{fig2} (a -b) Emulsion structures in microchannels. The different patterns are obtained by tuning the dispersed-to-continuous inlet flow ratios from $\phi=u_d/2u_c=1 \div 3.6$. Simulations are compared aginst the experiments of Marmottant et al. \cite{marmottantprl2009}. (a) Hex-Three, (b) Wet Hex-Two, (c) Dry Hex-Two, (d) Hex-One. }
\end{center}
\end{figure}
\begin{figure}
\begin{center}
\includegraphics[scale=1.0]{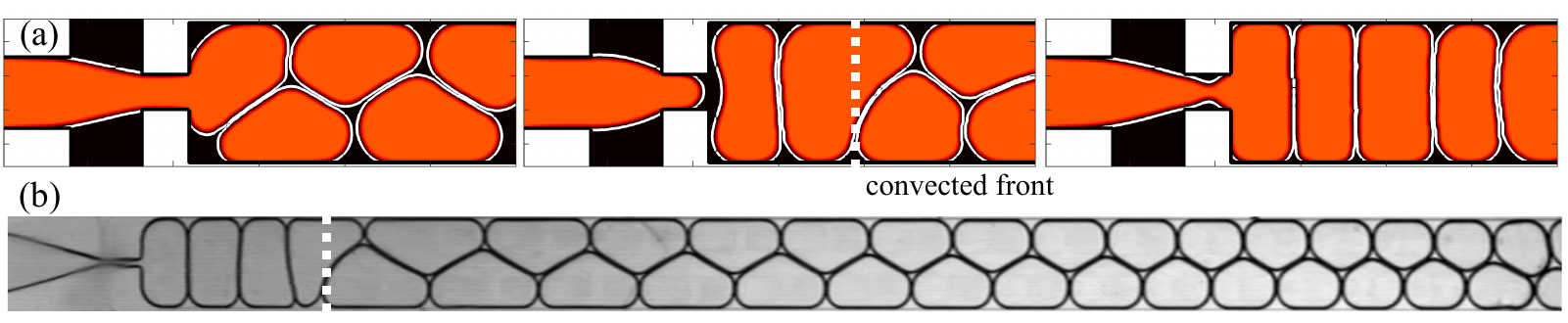}
\caption{\label{fig3} (a -b) Spontaneous dynamic rearrangement of the emulsion pattern within the outlet micro-channel ((a) numerical simulations, (b) experiments). The vertical bars indicate the rearrangement fronts. Lower panel, experimental snapshot taken from \cite{marmottantprl2009} }
\end{center}
\end{figure}
The droplet production can be controlled by tuning the dispersed to continuous flow rate ratio
and, when the density of droplets is sufficiently high, an emulsion is formed within the exit micro-channel.
As reported in figure \ref{fig2}, different crystalline structures appear when lowering the inlet dispersed 
fluid discharge for a constant inlet continuous phase flow rate:
the droplets arrangement within the microchannel transits from a Hex-three structure, typical of flowing emulsions in microchannels, to 
wet and dry Hex-two structures, with the droplets self-arranging in a ordered two files configuration, up to the 
"bamboo" structure typical of Hex-one-like emulsions.
Remarkably, the numerical model employed, is able to seamlessly and accurately capture all these regimes, as evidenced in 
figure \ref{fig2}, in which the pattern obtained by means of numerical simulations are compared with those reported in \cite{marmottant2009,marmottantprl2009}.

Between these homogeneous crystal patterns, we also observed a self-regulated transition flow, in agreement with \cite{marmottantprl2009}, as reported
in figure \ref{fig3}. This intermediate regime is associated with a rich dynamic behaviour, resulting in structures that vary over space and time
within the channel and controlled by the speed of the foam.

In particular, we observe a spontaneous transition regime in which the Hex-two and the Hex-one
structures simultaneously coexist. This regime is characterized by the presence of a rearrangement front 
travelling within the outlet channel along the flow direction with a speed smaller than that of the foam (see \cite{marmottant2009}).
As observed in \cite{marmottantprl2009}, this meta-stable regime strikingly resembles a thermodynamic first order transition.\\
Moreover, we inspected the pressure field within the inlet channel of the dispersed component, noting that the transition between the two crystal patterns is signaled by a rapid increase of the pressure of the dispersed phase, which is a feedback effect of the partial clogging of the outlet channel caused by the droplets arranged in a two files Hex-two crystal. The pressure undulations are arguably due to the periodic pinch-off of the droplet within the orifice and its subsequent injection in the outlet channel.
This complex flow is a genuine example of dynamic self-organization in microfluidic channels which prospects new chances for the generation of highly controlled emulsions.

In particular, this intermediate regime offers a route to the production of droplets samples with a controlled polydispersity.
\begin{figure}
\begin{center}
\includegraphics[scale=0.7]{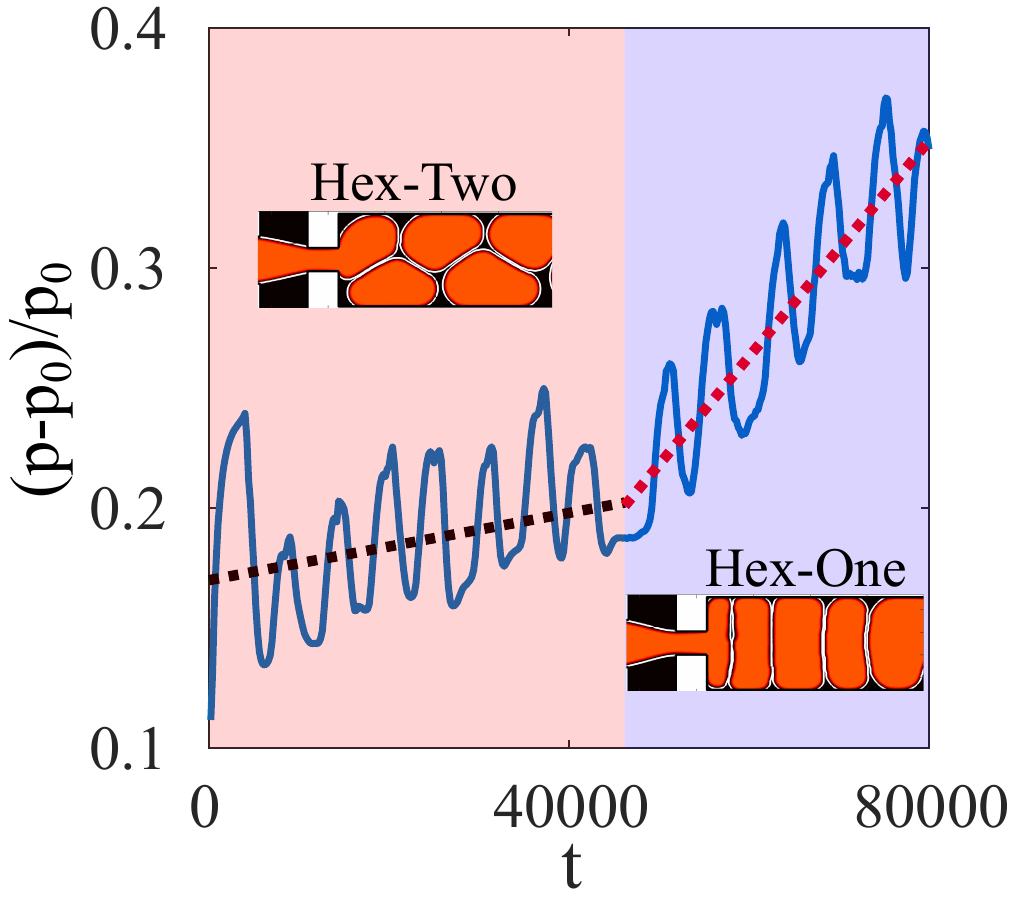}
\caption{\label{fig4} Non dimensional pressure variation ($p_0$ is a reference bulk pressure) within the inlet main channel of the dispersed phase as a function of time. The dotted lines are linear fits to the data. The transition from an Hex-Two to an Hex-One crystal is clearly indicated by a sudden jump in the slopes of the linear fits. }
\end{center}
\end{figure}

\section{Conclusions and future outlook}

Summarising, we have presented a new lattice Boltzmann method 
for multicomponent fluids, augmented with near-contact repulsion between 
approaching droplets, due to the presence of surfactant molecules.

The near-contact interactions are represented by coarse-grained effective forces
which, besides preventing droplet coalescence, prove capable of predicting non-trivial
large-scale features of the flow, such as the formation of regular configurations of flowing
droplets, known as soft flowing crystals.

The success of the method crucially hinges on the universality of the underlying physics; in other
words, the key requirement on near-contact interactions is to prevent droplet coalescence, and once
this feature is secured, the highly non-trivial dynamics which develops at larger space and time
scales goes under control of other mechanisms, such as capillarity and hydrodynamic interactions, which 
are also  taken in charge by the mesoscale lattice Boltzmann method.

Whilst based on a dramatic simplification of the underlying physics at the molecular level, the results obtained in this paper suggest that, at least at the spatial scale at hand, a coarse grained description is appropriate to describe the  mesoscale evolution of an interacting multidroplet system.

The present work is expected to benefit the multiscale simulation of microfluidic 
devices for the production of new droplet/bubble-based mesoscale porous materials.

\section{Acknowledgments}
A. M., M. L., A. T. and S. S. acknowledge funding from the European Research Council under the European Union's Horizon 2020 Framework
Programme (No. FP/2014-2020) ERC Grant Agreement No.739964 (COPMAT).

\end{document}